\documentclass[%
 reprint,
 amsmath,amssymb,
 aps,
]{revtex4-1}

\usepackage{graphicx}% Include figure files
\usepackage{dcolumn}% Align table columns on decimal point
\usepackage{bm}
\usepackage[dvipsnames]{xcolor}

\begin{document}

\preprint{APS/123-QED}

\title{Realizing split-pulse x-ray photon correlation spectroscopy to measure ultrafast dynamics in complex matter}% Force line breaks with \\

\author{Yanwen Sun}
\altaffiliation[Also at ]{Physics Department, Stanford University, Stanford, California, 94305, USA}
\affiliation{Linac Coherent Light Source, SLAC National Accelerator Laboratory, Menlo Park, California, 94025, USA}
\author{Mike Dunne}
\affiliation{Linac Coherent Light Source, SLAC National Accelerator Laboratory, Menlo Park, California, 94025, USA}
\author{Paul Fuoss}
\affiliation{Linac Coherent Light Source, SLAC National Accelerator Laboratory, Menlo Park, California, 94025, USA}
\author{Taito Osaka}
\affiliation{RIKEN SPring-8 Center, Sayo, Hyogo, 679-5148, Japan}
\author{Aymeric Robert}
\affiliation{Linac Coherent Light Source, SLAC National Accelerator Laboratory, Menlo Park, California, 94025, USA}
\author{Mark Sutton}
\altaffiliation[Also at ]{Linac Coherent Light Source, SLAC National Accelerator Laboratory, Menlo Park, California, 94025, USA}
\affiliation{Physics Department, McGill University, Montr\'eal, Quebec, H3A 2T8, Canada}
\author{Makina Yabashi}
\affiliation{RIKEN SPring-8 Center, Sayo, Hyogo, 679-5148, Japan}
\author{Diling Zhu}
\email{dlzhu@slac.stanford.edu}
\affiliation{Linac Coherent Light Source, SLAC National Accelerator Laboratory, Menlo Park, California, 94025, USA}

\date{\today}% It is always \today, today,

\begin{abstract}
Split-pulse x-ray photon correlation spectroscopy has been proposed as one of the unique capabilities made possible with the x-ray free electron lasers. It enables characterization of atomic scale structural dynamics that dictates the macroscopic properties of various disordered material systems. Central to the experimental concept are x-ray optics that are capable of splitting individual coherent femtosecond x-ray pulse into two distinct pulses, introduce an adjustable time delay between them, and then recombine the two pulses at the sample position such that they generate two coherent scattering patterns in rapid succession. Recent developments in such optics showed that, while true `amplitude splitting' optics at hard x-ray wavelengths remains a technical challenge, wavefront and wavelength splitting are both feasible, able to deliver two micron sized focused beams to the sample with sufficient relative stability. Here, we however show that the conventional approach to speckle visibility spectroscopy using these beam splitting techniques can be problematic, even leading to a decoupling of speckle visibility and material dynamics. In response, we discuss the details of the experimental approaches and data analysis protocols for addressing issues caused by subtle beam dissimilarities for both wavefront and wavelength splitting setups. We also show that in some scattering geometries, the $Q$-space mismatch can be resolved by using two beams of slightly different incidence angle and slightly different wavelengths at the same time. Instead of measuring the visibility of weak speckle patterns, the time correlation in sample structure is encoded in the `side band' of the spatial autocorrelation of the summed speckle patterns, and can be retrieved straightforwardly from the experimental data. We demonstrate this with a numerical simulation.
\end{abstract}

\maketitle

%\tableofcontents
\section{Introduction}
\begin{figure}[ht]
\centering
\includegraphics[width=\linewidth]{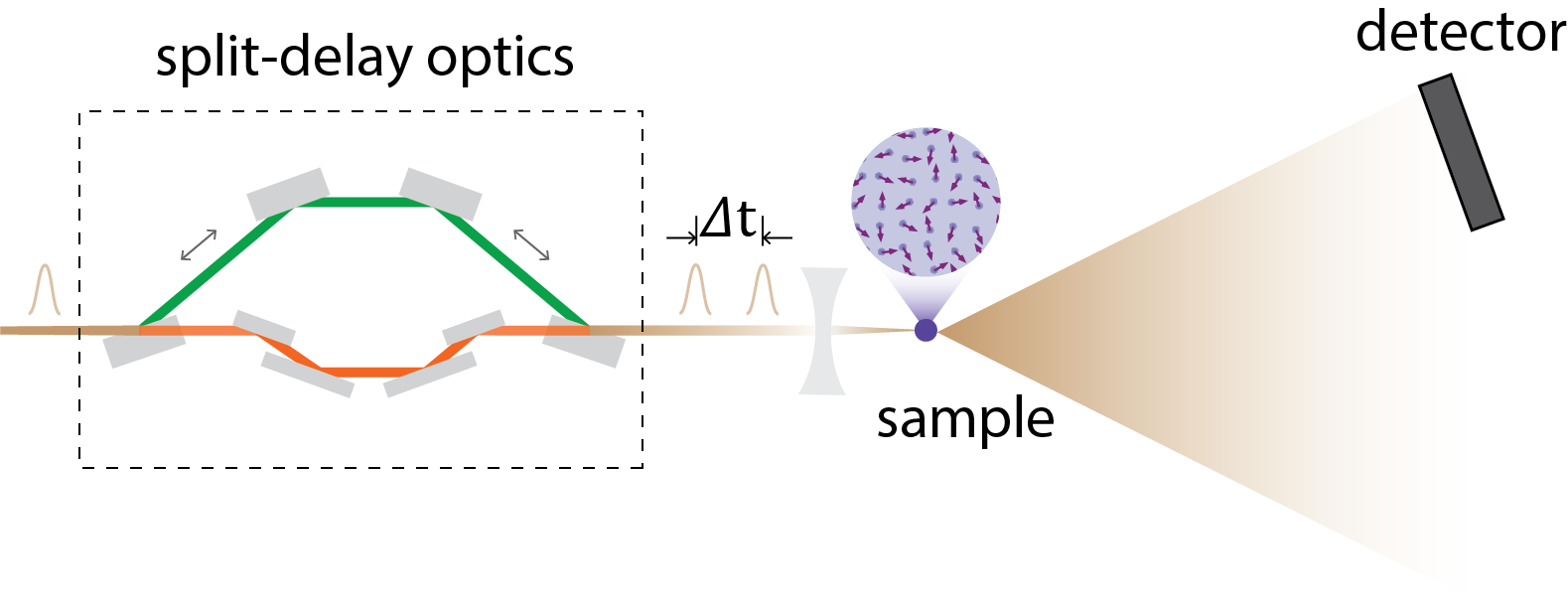}
\caption{Schematics of a generic split-pulse XPCS experiment using crystal optics based hard x-ray split-delay. A set of crystals are arranged such that individual pulses are split, delayed in time as compared to each other and subsequently focused and recombined at the sample location. Arrows along the green beam path indicate how this path length can be adjusted by moving some of the crystals within the split-delay optics. The summed coherent scattering patterns measured for each pulse pair is recorded by an x-ray imaging detector located downstream the sample at a given scattering angle.}
\label{fig:fig1}
\end{figure}

Nearly fully transversely coherent femtosecond x-ray pulses produced by x-ray free electron laser (FEL) sources opened up the possibilities of direct measurement of atomic scale dynamics of complex systems at their native time scales~\cite{stephenson2009x}. One area of particular interest is the investigation of noncrystalline matter such as liquids, glasses, amorphous and disordered systems, and holds the promises of unlocking the mysteries behind the glass transition, liquid-liquid phase transitions, fragile-to-strong transitions, to name a few~\cite{berthier2011theoretical,shintani2008universal,langer2008anomalous}. A primary methodology with the potential to extend dynamic light scattering to angstrom and femto-/picosecond time scale is the so-called split-pulse x-ray photon correlation spectroscopy (XPCS) technique, where the dynamics of the scattering object are imprinted onto the fluctuations of coherent scattering intensity distribution~\cite{GRUBEL_SDXPCS_2007}. The schematic of a generic split-pulse XPCS experiment is illustrated in Fig.~\ref{fig:fig1}. Two delayed beams are generated by a split-delay optics, and then focused down to a small size at the sample location. Downstream the sample a pixelated X-ray detector measures coherent scattering patterns. While area detectors capable of independently measuring the scattering patterns from two subsequent x-ray pulses with a femto- to picosecond separation will not be available in the foreseeable future, it was proposed that the correlations between the coherent scattering patterns from the two successive pulses can nevertheless be obtained from the summed scattering pattern, by analysing the speckle visibility~\cite{shenoy2000lcls}. The dependence of the visibility, as a function of of the temporal separation between the two pulses, thus carries the potential to provide detailed information on the dynamics information of the system being probed~\cite{gutt2009measuring}.

The purpose of x-ray split-delay lines is to generate x-ray pulse pairs with continuously adjustable time separations in the femto- and picosecond time range. The generic split-delay-recombine optical arrangement has been realized recently with increased robustness, primarily in the form of wavefront or wavelength splitting setups. Pulse pairs can now routinely be generated and delivered to a sample with sufficient reliability and stability for two pulse coherent scattering measurements~\cite{Osaka2017,Shi2018,sun2019compact}. In this work, we present detailed examinations of the speckle correlation analysis in these scenarios, illustrate the incompatibility of the wavefront and wavelength splitting optical schemes with the speckle visibility spectroscopy concept.  We propose an alternative correlation extraction methodology, as well as a $Q$-space compensation solution by using two different wavelengths, that allows the extraction of dynamics under the general experimental scheme of two-pulse XPCS. We also discuss optimization of real experiment parameters.
\section{Split-pulse scattering geometry}
The two-pulse XPCS measurement concept envisioned the use of two identical x-ray pulses, i.e, having the same photon energy, trajectory, beam profile, wavefront, and coherence properties, with an adjustable time separation. This was initially proposed to be realized by using thin crystal optics with thickness smaller than the extinction depth of the chosen x-ray Bragg reflection~\cite{roseker2009performance}. However, the fabrication and handling of sufficiently thin and robust beam splitting crystals still remains to date a major technical challenge. 

Two alternative splitting techniques, wavelength and wavefront splitting, have been adopted during the past few years in Bragg crystal based x-ray split-delay optics~\cite{roseker2009performance,Osaka2016,Osaka2017,zhu2017development,sun2019compact}. While these systems have shown great progress towards delivering two similar x-ray foci to the sample with fine control of their time delay and good relative beam position stability, these splitting techniques lead to other ineluctable differences in the two beams/pulses. For example, Roseker~\emph{et al.} and Osaka~\emph{et al.} used thin silicon crystals as beam splitters~\cite{roseker2009performance,Osaka2016}. However, the available thin crystals are still thicker than the extinction depth of the reflection. As a result, the portion of wavelengths that falls within the reflecting bandwidth gets almost fully reflected, while the other wavelengths transmit through the crystal. The two output beams as a result will have different photon energies.
More recent x-ray split-delay optics adopted the wavefront splitting geometry~\cite{Osaka2017,zhu2017development,sun2019compact}: part of the incoming beam hits a polished edge of the beam splitting crystal, meets Bragg condition and gets reflected, while the other part of the beam passes over the edge. The split beams are directed into different beam paths within the split-delay optics before getting recombined using another crystal with a polished edge. In this case, the two parts of the recombined beam are parallel but not exactly collinear. Experimentally, when trying to bring the both parts of the beam to the same location on the sample with focusing optics, there will be an inevitable crossing angle between the two beams. 

The slight differences in the two `probe' beam properties will lead to a mismatch in their scattering in the far field, which could in principle compromise our ability to recover the desired material dynamics. Below we provide a generalization of this mismatch originating from those differences.

\begin{figure}[h!]
\centering
\includegraphics[width=0.8\linewidth]{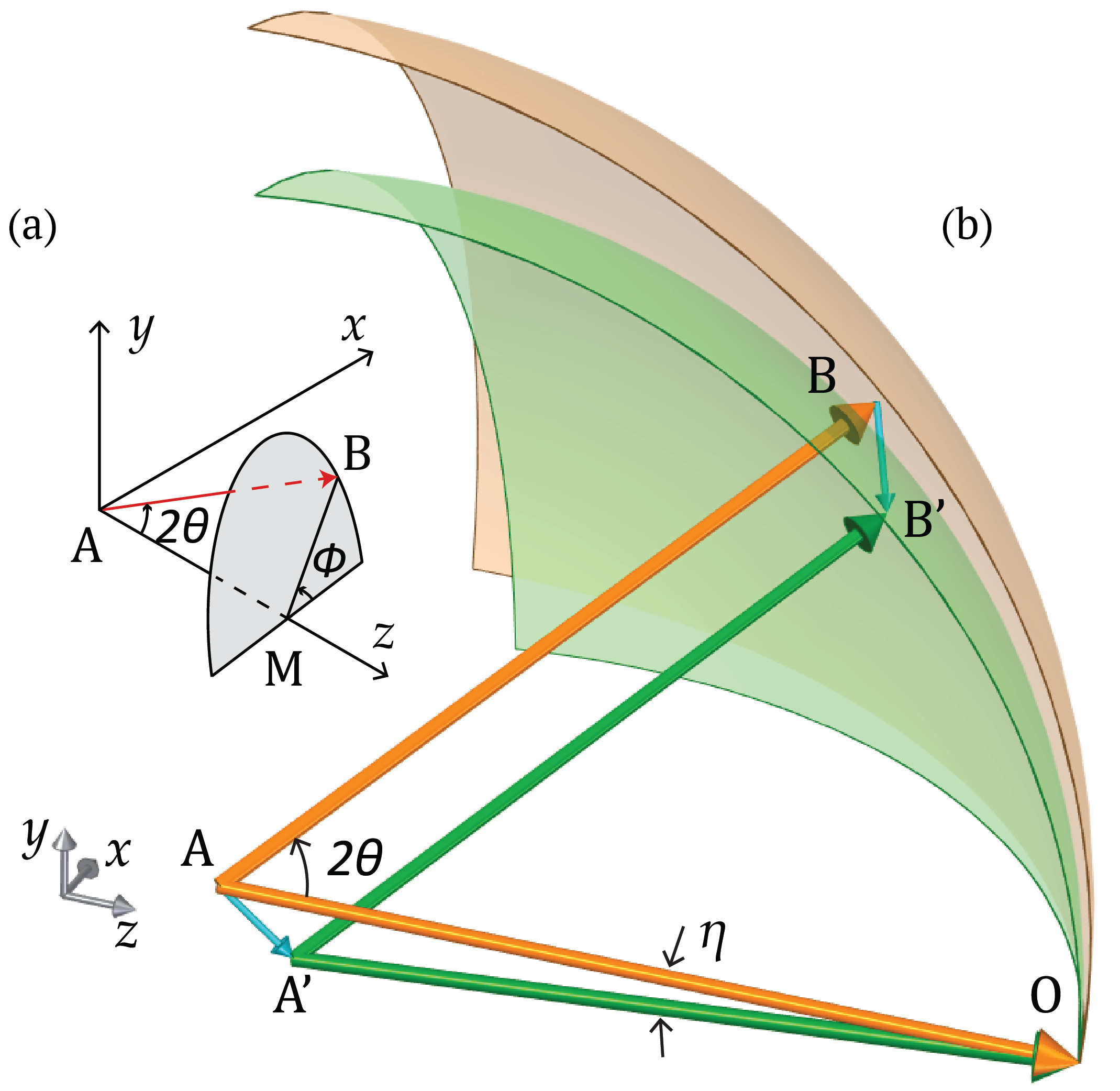}
 \caption{Illustration of the Ewald spheres considering the differences between the two recombined beams.(a) Coordinate system definition (b) Illustration of the mismatch of scattering vectors in the reciprocal space. The two beams are denoted in orange and green. They can either have slight different wavelengths corresponding to the radius change of the Ewald sphere or different incident angles corresponding to the rotation of the Ewald sphere. }
\label{fig:Ewald}
\end{figure}
As shown in Fig.~\ref{fig:Ewald}(a), we define $z$ as the incident beam propagation direction and $y$ as the direction along which their trajectories deviate from one another. Using the exit beam wavevector $\bm{k_f}$, we can define the spherical coordinate: $2\theta$ as the angle with respect to $z$ axis, $\phi$ as the angle of its projection on $xy$ plane (BM) and $x$ axis ($0\le \phi < 2\pi$). The scattering experiment can be presented as shown in Fig.~\ref{fig:Ewald}(b) in the reciprocal space: with two Ewald spheres denoted in orange and green for the two output beams. Their two slightly different radius $k_i$ and $k_i'$ represent the difference between their photon energies. Its length is thus related to the difference in the wavelength $\delta \lambda$: 
\begin{equation}
    \delta k_i = k_i' - k_i  = k_f' - k_f =  \frac{\delta \lambda}{\lambda}k_i.
    \label{eq:lambdak}
\end{equation}
$\bm{k_i} = \bm{\mathrm{AO}}$ and $\bm{k_i'} = \bm{\mathrm{A'O}}$ are the incidence wavevectors. $\eta$ is the angle between the two indicating their slight different incident angle on sample. We use $\bm{k_f} = \bm{\mathrm{AB}}$ and $\bm{k_f'}=\bm{\mathrm{A'B'}}$ for the two output wavevectors. A chosen detector pixel can be represented by the parallel exit wavevectors $\bm{k_f}\parallel \bm{k_f'}$ for the two beams respectively. In the Cartesian coordinate system defined by $x,y,z$, the incidence and exit wavevectors for both beams can be written as:
\begin{equation}
\begin{split}
    \bm{k_i} &= \bm{\mathrm{AO}} = k_i[0,0,1],\\
    \bm{k_i'} &=\bm{\mathrm{A'O}}  = k_i' [0,\sin\eta,\cos\eta],\\
    \bm{k_f} & = k_i[\sin2\theta \cos\phi,\sin2\theta \sin\phi,\cos2\theta],\\
    \bm{k_f'} & = k_i'[\sin2\theta \cos\phi,\sin2\theta \sin\phi,\cos2\theta].
    \end{split}
\end{equation}
The difference in the momentum transfer at the same detector pixel location $\bm{\mathrm{BB'}} = \bm{\mathrm{OB'}}-\bm{\mathrm{OB}}$ can be derived as:
\begin{equation}
    \begin{split}
        \bm{\mathrm{BB'}} &=  (\bm{k_f'}-\bm{k_i'}) - (\bm{k_f}-\bm{k_i})\\
         & = (k_i'-k_i)[\sin2\theta \cos\phi,\sin2\theta \sin\phi,\cos2\theta]\\
        &+[0,-k_i'\sin\eta,k_i-k_i'\cos\eta].
    \end{split}
\end{equation}
An area detector samples the speckles that lie on the two Ewald spheres separately for the two beams. $\bm{\mathrm{BB'}}$ is a measure of the deviation of the momentum transfer $Q$ measured by the same detector pixel. 

We next discuss this $Q$ mismatch for wavefront and wavelength splitting schemes respectively and the resulting constraints on the experimental geometry and sample parameters. For the rest of the paper, we will choose a photon energy of 10~keV, and a bandwidth $\delta \lambda /\lambda = 5.6 \times 10^{-5}$ (FWHM) and momentum transfer of interest at $Q = 2~\mathrm{\AA^{-1}}$ corresponding to  $\theta \approx 11.38^{\circ}$ for experimental case studies. A few assumptions are made for speckle size calculation and we follow the methods explained in details in Ref.~\cite{mark2019,sandy1999design,lumma2000area}.
\section{Wavefront splitting case}
\begin{figure}[h!]
\centering
\includegraphics[width=0.9\linewidth]{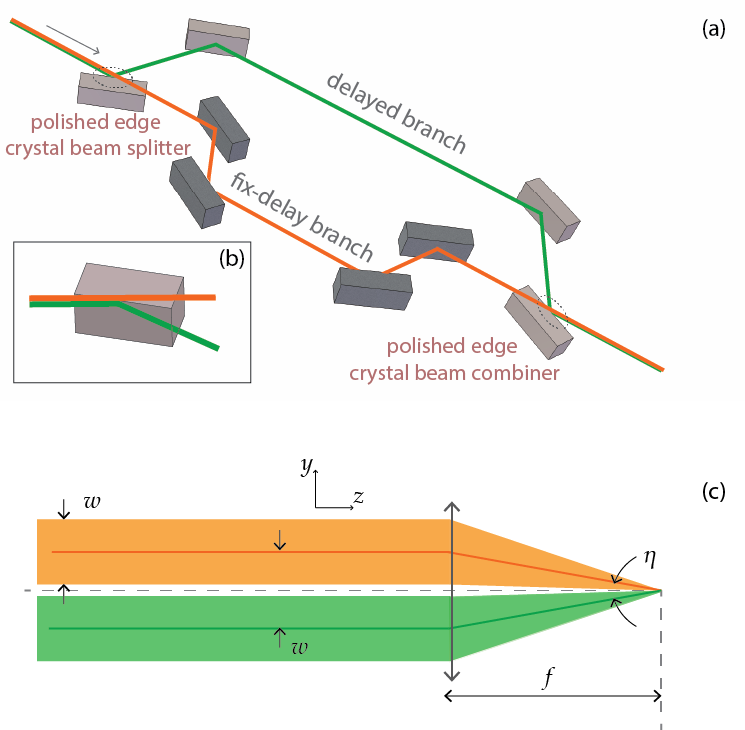}
 \caption{(a) Optical arrangement of a split-delay system based on wavefront splitting using crystals with polished edges. (b) illustration of the wavefront splitting/combining process as indicated by the dashed oval in (a). (c) Illustration of the crossing angle between the two beams after focusing optics due to the non collinear geometry.}
\label{fig:crossing}
\end{figure}

In this section we discuss the case of wavefront splitting. Figure~\ref{fig:crossing}(a) is a schematic of the realization of the split-delay based on polished edge crystals. After beam recombination at the crystal beam combiner, the two output beams travel nearly collinearly in order to achieve spatial overlap at the sample using focusing optics, as illustrated in Fig.~\ref{fig:crossing}(c). The magnitude of the minimum crossing angle $\eta$ is therefore determined by the beam width $w$ (defined as in Fig.~\ref{fig:crossing}(c)) of the unfocused beam and the focal length $f$:
\begin{equation}
    \eta \approx \frac{w}{f}.
\end{equation}
We also assume the two output beams have the same photon energy, and thus $k_i = k_i'$. The $Q$ space mismatch then reduces to $\bm{\mathrm{BB'}}= [0,-k_i\sin\eta,k_i(1-\cos\eta)]$. Its magnitude $\mathrm{BB'} = 2k_i\sin(\eta/2)$ is invariant of $\theta$ and $\phi$. $\bm{\mathrm{BB'}}$ can be decomposed into its \textit{in}- and \textit{out-of-detector-plane} components. In order for the same pixel to be mapped to the same speckle, the magnitude of the out-of-detector-plane mismatch $\bm{\mathrm{BC}}$ should be much smaller than the speckle ellipsoid size along the exit wavevector direction ($\bm{k_f}$ or $\bm{k_f'}$). Otherwise the detector will be sampling a completely different slice of the 3D $Q$ space. The two speckle patterns will have no correlation as a result. The out-of-detector-plane mismatch can be written as
\begin{equation}
\begin{split}{}
     \mathrm{BC} &=  \bm{\mathrm{{BB'}}}\cdot \frac{\bm{k_f}}{k_f} \\ &= k_i[-\sin2\theta \sin\phi\sin\eta+\cos2\theta(1-\cos\eta)] \\
     & = -k_i\eta \sin 2\theta \sin \phi  + k_i O(\eta^2).
     \label{eq:shift_wavefront}
     \end{split}
\end{equation}
Here we denote the sum of all higher order terms of $\eta$ as $O(\eta^2)$ because typically $\eta$ is on the order of $10^{-4}$ considering the small numerical aperture of the x-ray focusing optics. It has a dependence on both $2\theta$ and $\phi$. To first order, the magnitude of BC is maximum for $\phi = \pi/2$ and will be minimized for $\phi = 0$ where it is $k_iO(\eta^2)$. 

Similarly, the in-detector-plane mismatch:
\begin{widetext}
$$     \bm{\mathrm{CB'}}=  \bm{\mathrm{BB'}} - \bm{\mathrm{BC}} = k_i[\frac{1}{2}\eta\sin^22\theta\sin 2\phi+O(\eta^2), (-1+\sin^2 2\theta \sin^2\phi)\eta+O(\eta^2), \frac{1}{2}\eta\sin 4\theta \sin \phi +O(\eta^2)].
$$
\end{widetext}
Its length
\begin{equation}
    \mathrm{CB'} = k_i\eta\sqrt{1-\sin^22\theta\sin^2\phi}+O(\eta^2)
\end{equation}
reaches a maximum of $k_i \eta$ for $\phi = 0$ and a minimum of $k_i \eta \cos2\theta$ for $\phi = \pi/2$. The direction of in-detector-plane mismatch is $\bm{s} = [0,1,0]$ for $\phi = 0$ and $\bm{s} =  [0,-\cos2\theta, \sin2\theta]$ for $\phi = \pi/2$. 

\begin{figure}[h!]
\centering
\includegraphics[width=0.9\linewidth]{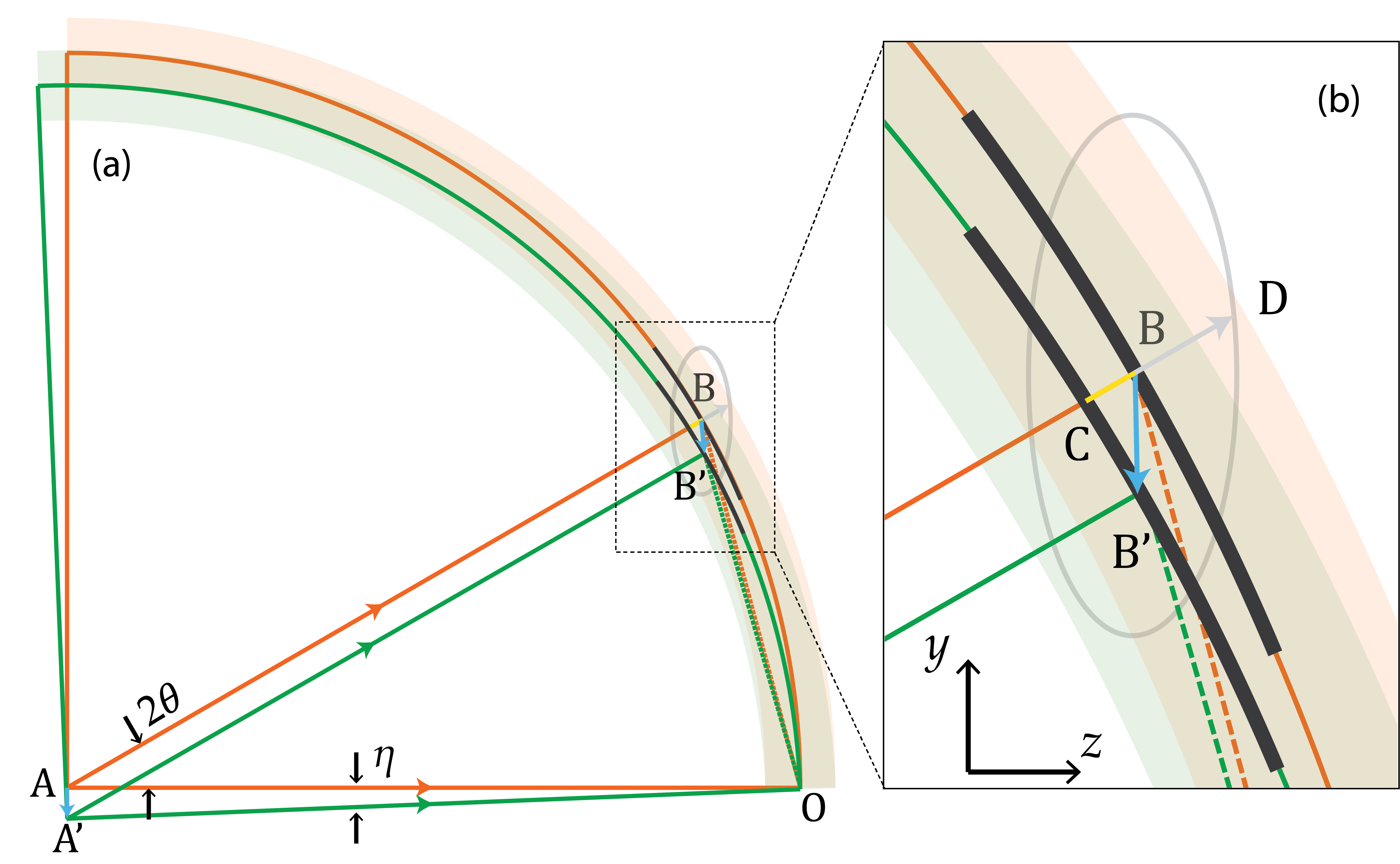}
 \caption{Illustration of wavefront splitting for $\phi = \pi/2$. (a) Reciprocal space illustration of the scattering of two pulses from split-delay optics using wavefront splitting and recombining. The two pulses are plotted in orange and green and have a crossing angle $\eta$ between the incident wavevector $\bm{k_i}=$\textbf{AO} and $\bm{k_i'}=$\textbf{A'O}. The incident beam bandwidth is indicated by the thickness of the Ewald circle in shade orange/green. At the same detector location ($2\theta$ with respect to $\bm{k_i}$), the measured wavevectors are respectively \textbf{OB} and \textbf{OB'}. (b) A zoomed in view of the rectangular area in (a). The blue arrow represents the momentum transfer mismatch $\bm{\mathrm{BB'}}$. A speckle ellipsoid is plotted in light gray and the detector locations are plotted in black for the two Ewald circles.}
\label{fig:wavefront}
\end{figure}
The speckle size, on the other hand, is determined to the first order by the focal spot size and the sample thickness $t$. The largest possible speckle size is reached at the diffraction limited focal spot size of $w_0$:  
\begin{equation}
    w_0 \approx \frac{4}{\pi} \frac{ \lambda }{w}{f}.
    \label{eq:focalspot}
\end{equation}
Following Ref.~\cite{sikorski2015focus}, the rms speckle size in the $y$ direction $S_y \approx 0.38 k_i\lambda/w_0 \approx 0.30 k_i \eta$. Define $c$ as the ratio between the rms speckle size $S_y$ and the scattering mismatch $\mathrm{BB'}\approx k_i\eta$, i.e.,
\begin{equation}
    c = \frac{\alpha S_y}{k_i \eta},
\end{equation}
where $\alpha = \sqrt{6}$ is to convert the mismatch to rms and is explained in detail in Ref.~\cite{lumma2000area}. For $\phi = 0$, along $\bm{s}$, the rms speckle size $S_{\bm{s}} = S_y$, while the in-detector-plane mismatch is $k_i\eta$. $c \approx 0.73$ suggests that the $Q$ space mismatch in the detector plane is generally larger than one speckle size. This leads to the unavoidable reduction in contrast in the sum-speckle pattern. As a result, visibility analysis will become significantly less sensitive to sample dynamics.

For $\phi = \pi/2$ as illustrated in Fig~\ref{fig:wavefront}, along $\bm{s}$, the rms speckle size is
\begin{align}
    S_{\bm{s}} &= \frac{S_yS_z}{\sqrt{S_z^2\cos^22\theta+S_y^2\sin^22\theta}},
\end{align}
where $S_z$ is the rms speckle size along $z$, which is to first order determined by the thickness of the sample ~\cite{hruszkewycz2012high} $$S_z \sim \frac{k_i\lambda}{t}.$$
(Note that longitudinal coherence also plays a role. For computing $S_z$, we follow the numerical methods provided in Ref.~\cite{mark2019}.) Its ratio with respect to CB' is
\begin{equation}
    \begin{split}
        \frac{\alpha S_{\bm{s}}}{\mathrm{CB'}} &\approx  \frac{c}{{\cos2\theta\sqrt{\cos^22\theta +\frac{S_y^2}{S_z^2}\sin^2 2\theta}}} \\
        &< \frac{c}{\cos^2 2\theta} \approx 0.86.
    \end{split}
\end{equation}
Just as the $\phi = 0$ case, here the in-detector-plane mismatch is inevitably larger than the speckle size. We will provide an analytical solution to address the in-detector-plane mismatch in later sections. However, the out-of-plane mismatch will have to be minimized. Ideally, BD, the out-of-detector-plane speckle size defined in Fig~\ref{fig:wavefront}(b), shall be much larger than the out-of-plane $Q$ mismatch, i.e.,
\begin{equation}
    \frac{2\mathrm{BD}}{\mathrm{BC}}\gg 1.
\end{equation}
For our case study, at $\phi=\pi/2$, following
$$\mathrm{BD}= \frac{\alpha S_yS_z}{2\sqrt{S_z^2\sin^22\theta+S_y^2\cos^22\theta}},$$

we have
\begin{equation}
    \begin{split}
        \frac{2\mathrm{BD}}{\mathrm{BC}}
        &= \frac{c}{\sin 2\theta \sqrt{\sin^2 2\theta +\frac{S_y^2}{S_z^2}\cos^2 2\theta}}
        \\&<\frac{c}{\sin^2 2\theta}\approx 4.9,
    \end{split}
\end{equation}
when $S_y<<S_z$, or when the sample thickness is much smaller than $w_0$. Typical values of $w = 100~\mu m$, $f = 1$~meter give $\eta = 10^{-4}$, $\mathrm{BC} \approx 2.0 \times 10^{-4} \mathrm{\AA^{-1}}$, $w_0\approx 1.6~\mu m$ and $S_y \approx 1.5 \times 10^{-4} \mathrm{\AA^{-1}}$. Experiments require to use very thin samples, $t\sim 100~nm$, which limits the total scattering signal.

For $\phi = 0$, considering a sample thickness of $t= 15~\mu m$, $S_z \approx 2.2 \times 10^{-5} \mathrm{\AA^{-1}}$, $\mathrm{BC} = k_i O(\eta^2)\sim 10^{-8} \mathrm{\AA^{-1}}$, we have $$ \frac{2\mathrm{BD}}{\mathrm{BC}} \approx 2.5\times 10^3\gg 1.$$ Clearly, in order to minimize the out-of-plane $Q$ mismatch, the $\phi=0$ configuration would be more advantageous than $\phi=\pi/2$. However, the in-detector-plane $Q$ mismatch would still make speckle visibility spectroscopy infeasible. Alternative correlation extraction method will be discussed in a later section. 
\section{Wavelength splitting case}
\begin{figure}[h!]
\centering
\includegraphics[width=\linewidth]{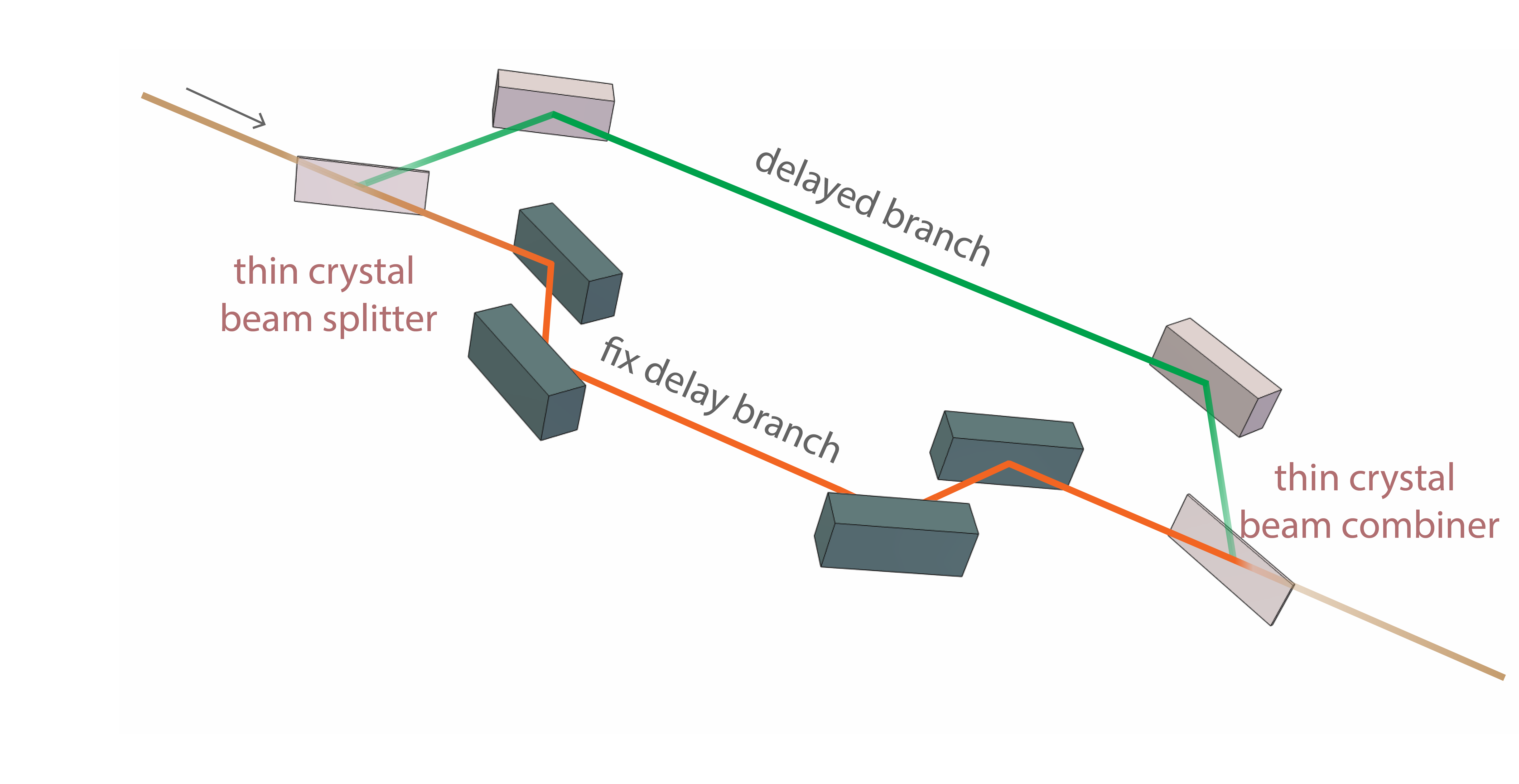}
 \caption{Optical arrangement of a split-delay system using thin crystal wavelength splitting.}
 \label{fig:splitdelay_wavelength}
 \end{figure}
\begin{figure}[h!]
\centering
\includegraphics[width=0.9\linewidth]{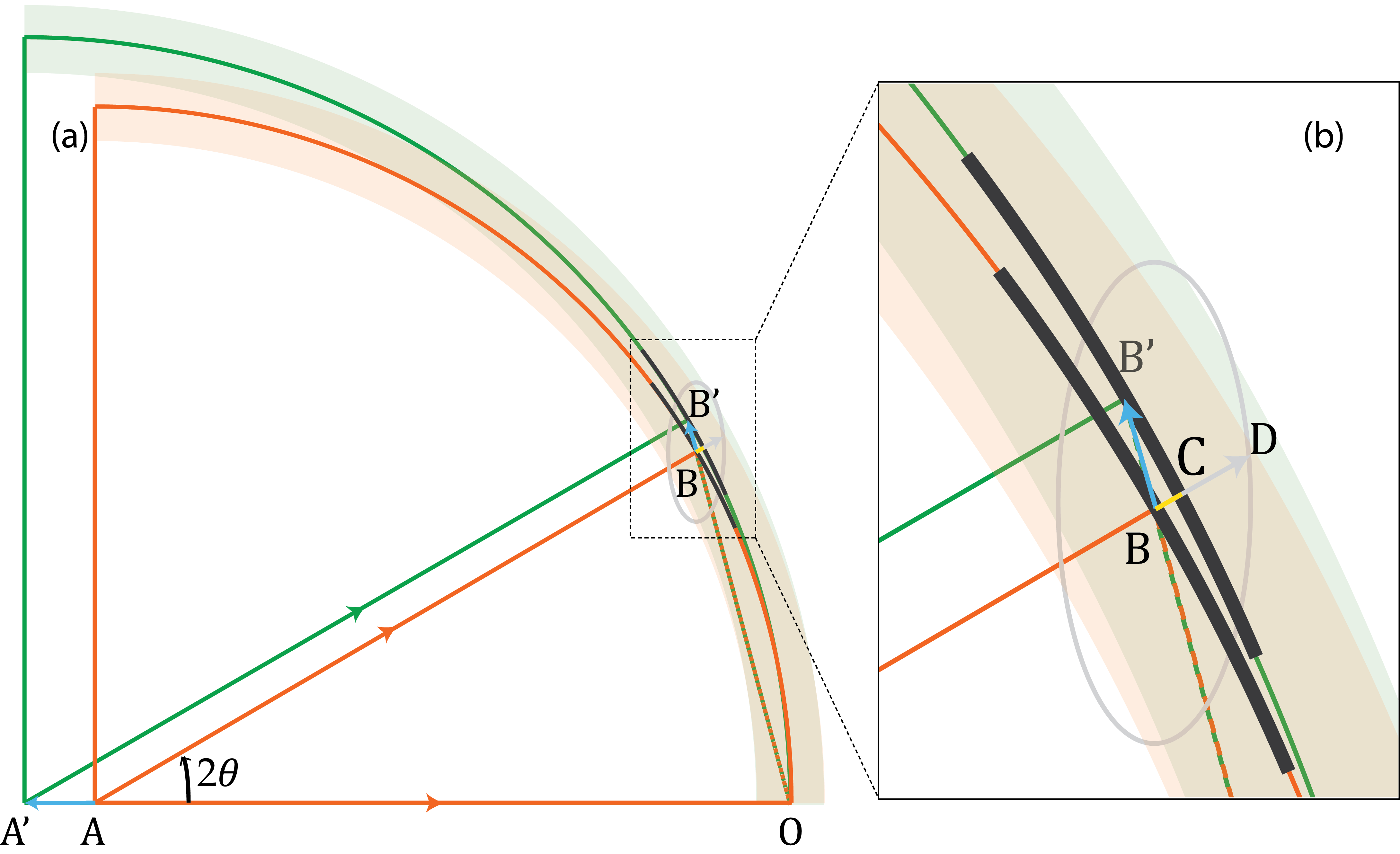}
\caption{Illustration of wavelength splitting. (a) Reciprocal space illustration of the scattering of two pulses from split-delay optics using wavelength splitting and recombining. \textbf{OA} and \textbf{OA'} are incident beam wavevectors which have different magnitudes $\Delta k_i = \mathrm{AA'}$. At the same detector location of scattering angle $2\theta$, the measured wavevectors are respectively \textbf{OB} and \textbf{OB'}. Using this method, the two pulses have different center energies offset by the bandwidth of the crystal reflection. (b) A zoomed in view of the rectangular area in (a). A speckle ellipsoid is plotted in light gray and the detector locations are plotted in black for the two Ewald circles.}
\label{fig:wavelength}
\end{figure}

We now evaluate the wavelength splitting scenario. The schematics of this type of split-delay optics is illustrated in Fig.~\ref{fig:splitdelay_wavelength}. Similar to the wavefront splitting case, wavelength splitting also leads to a mismatch in $Q$ space sampling between the two beams, even though the two beams can be recombined with high degree of collinearity. This is because the magnitude of $\bm{k_i}$ and $\bm{k_i'}$ will be slightly different as a result of the wavelength difference. This is illustrated in Fig~\ref{fig:wavelength}. 
The $Q$ space mismatch $\bm{\mathrm{BB'}}$ can be written as:
\begin{equation}
    \bm{\mathrm{BB'}} = \delta k_i\,(\sin2\theta \cos\phi, \sin2\theta \sin\phi, -1 + \cos2\theta).
\end{equation}
Using Eq. \ref{eq:lambdak}, its length $\mathrm{BB'} = \delta \lambda/\lambda Q$. Here $Q = 2k_i\sin\theta$ is the momentum transfer for the $k_i$ (orange) and then $Q'= 2k_i'\sin\theta$ would be for the $k_i'$ (green). Similarly, we can derive the the `in' and `out' of detector plane mismatch:
\begin{equation}
    \mathrm{CB'} \sim \frac{\delta \lambda}{\lambda} Q \cos\theta
\end{equation}
\begin{equation}
    \mathrm{BC} \sim  \frac{\delta \lambda}{\lambda} Q \sin\theta
    \label{eq:shift_wavelength}
\end{equation}
Both in- and out-of-detector-plane mismatches have no dependence on $\phi$. Assuming $S_x = S_y$ in this case, the out-of-detector-plane speckle size BD and the in-detector-plane speckle size $S_{\bm{s}}$ are also only related to scattering angle $2\theta$. The ratios between the speckle sizes and the magnitude of the $Q$ mismatch are therefore:
$$
    \frac{\alpha S_{\bm{s}}}{\mathrm{CB'}} = \frac{\alpha S_yS_z} {Q \cos\theta \sqrt{S_z^2\cos^22\theta+S_y^2\sin^22\theta}}\cdot \frac{\lambda}{\delta \lambda},
$$
$$
    \frac{2\mathrm{BD}}{\mathrm{BC}}= \frac{\alpha S_yS_z}{Q\sin\theta\sqrt{S_z^2\sin^22\theta+S_y^2\cos^22\theta}}\cdot\frac{\lambda}{\delta\lambda}.
$$
With the same chosen experiment parameters provided earlier, $\mathrm{BC} \approx 2.2\times 10^{-5} \mathrm{\AA^{-1}}$ and $\mathrm{CB'} \approx 1.1\times 10^{-4} \mathrm{\AA^{-1}}$ for $\delta \lambda/\lambda = 5.6 \times 10^{-5}$. For a 15~$\mu m$ thick sample, $\mathrm{BD} \approx 2.9 \times 10^{-5} \mathrm{\AA^{-1}}$ and $S_{\bm{s}} \approx 5.3 \times 10^{-5} \mathrm{\AA^{-1}}$. The ratios of speckle size and the mismatch in and out of the detector plane are thus still of comparable magnitude, calculated to be $\alpha S_{\bm{s}}/\mathrm{CB'} \approx 1.2$ and $2\mathrm{BD}/\mathrm{BC} \approx 2.6$ respectively.

In order to make the mismatch sufficiently small compared to the corresponding speckle size for optimization of speckle visibility analysis, one can either reduce the illumination volume by the use of thin samples, or by using narrow x-ray bandwidths, e.g. 50-100 meV at 10 keV for our case study parameters. This will lead to additional x-ray pulse intensity fluctuations when operated under self amplified spontaneous emission (SASE) conditions, and calls for the development of improved stability and longitudinal coherence via seeding schemes~\cite{amann2012,inoue2019,CBXFEL}.

\section{$Q$-space compensation and correlation extraction}
\subsection{Compensation of the out-of-detector-plane mismatch}
\begin{figure}[h!]
\centering
\includegraphics[width=0.9\linewidth]{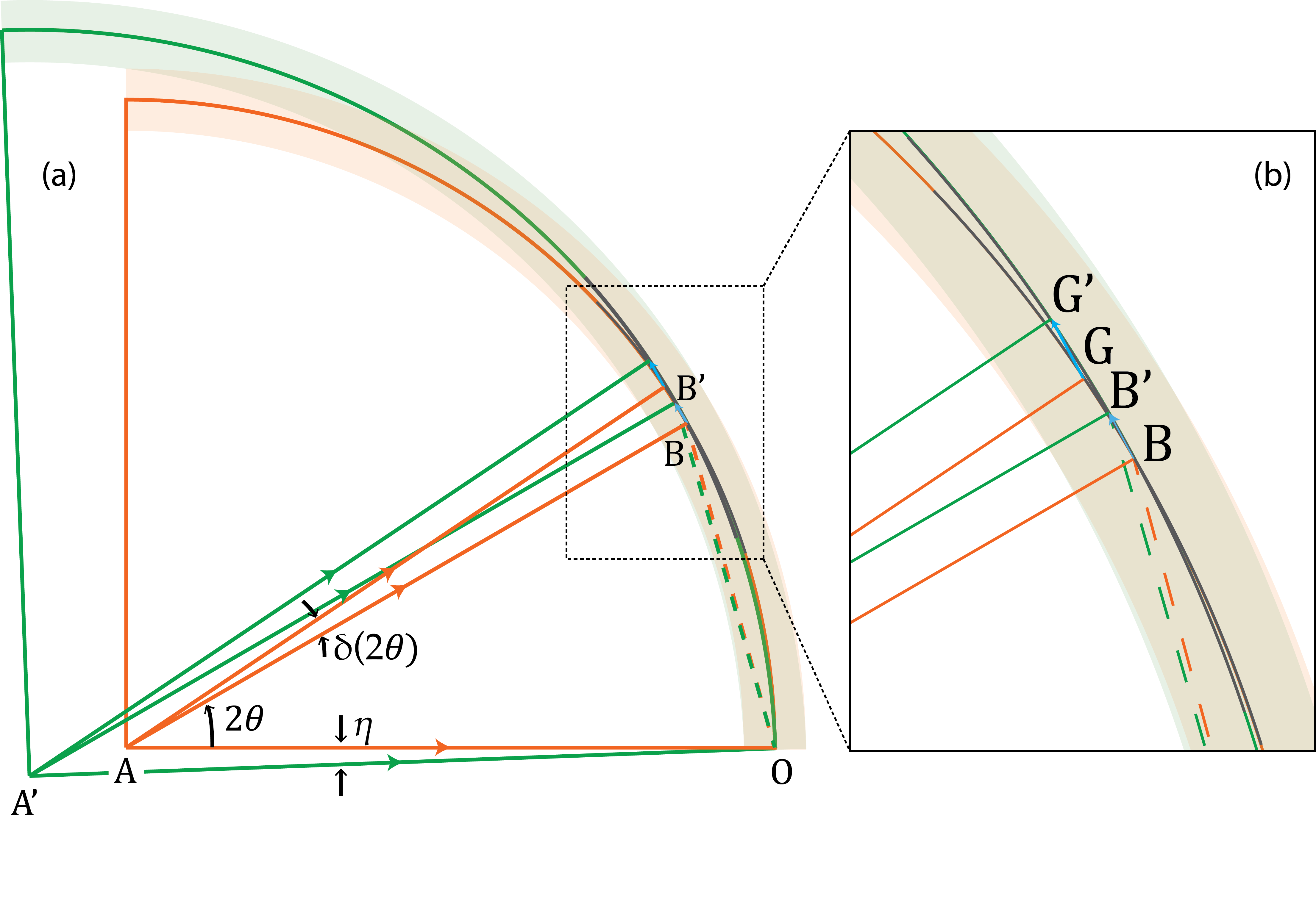}
\caption{ (a) Illustration of combing wavefront and wavelength splitting to miniize the out-of-detector-plane component of the $Q$ mismatch \textbf{BB'}. (b) A zoomed in view of the deviation between the Ewald spheres at $\bm{Q}$ denoted by \textbf{OG} after crossing at $\bm{Q}$ denoted by \textbf{OB}, we have $\angle \mathrm{GAB} = \delta (2
\theta)$ as the angle covered before the out-of-detector-plane mismatch is too large. }
\label{fig:solution}
\end{figure}

Following the formalism presented in the previous section, the out-of-detector-plane mismatch caused by the crossing angle between the two beams can be fully compensated by an intentional wavelength mismatch, as illustrated in Fig.~\ref{fig:solution}(a).
One could expand the green Ewald sphere around $O$ such that the two Ewald sphere cross each other again near B and B'. The goal is to have $\delta \bm{Q} =\bm{ Q} - \bm{Q'}$, or the vector \textbf{BB'}, in the tangential direction of the Ewald sphere. In this configuration, within a small $Q$ region near point $B$, the detector samples close-to-identical slices in the reciprocal space. In other words, \textbf{BB'} is perpendicular to \textbf{AB}, so in the triangular BOB', we have
$$
    \frac{\mathrm{OB'}}{\sin(180^{\circ}-\theta)} = \frac{\mathrm{OB}}{\sin(\theta-\eta/2)}.
$$
And this gives us
\begin{equation}
    \frac{\Delta \lambda}{\lambda} \sim \frac{\eta}{\tan\theta}.
\end{equation}
The same relationship can be obtained by equating the right side of Eq.~\ref{eq:shift_wavefront} and Eq.~\ref{eq:shift_wavelength}.
At 10~keV, use $Q = 2 \mathrm{\AA^{-1}}$ as the momentum transfer of interest, with $\eta = 10^{-4}$, we derive the required difference of the wavelength of the two split-delay branches to be $\Delta \lambda/\lambda\approx 5.0 \times 10^{-4}$, which is well within the SASE pulse bandwidth~\cite{bostedt2016linac}.

Another quantity we need to estimate is the scattering angle coverage $\delta \theta$, which is how large in scattering angle this method can correct before the out-of-detector-plane mismatch of momentum transfer becomes non-negligible, i.e., at point G and G' in Fig.~\ref{fig:solution}(b).
The deviation in the out-of-detector-plane direction is
\begin{equation}
    \begin{split}
        \delta \mathrm{BC} &\approx  \delta(\frac{\Delta \lambda}{\lambda} Q \sin\theta) -  \delta(\eta k_i\sin2\theta) \\
    &=2k_i\eta\delta \theta.
    \end{split}
\end{equation}
Using the parameters mentioned above, $\mathrm{BD} \approx 2.9 \times 10^{-5} \mathrm{\AA^{-1}}$. As $\eta = 10^{-4}$, $\delta \mathrm{BC} = \mathrm{BD}/2 \approx  1.5 \times 10^{-5}~\mathrm{\AA^{-1}}$ means $\delta (2\theta) \approx   0.029$ (or $\sim$144~mm at 5 meter detector distance). This can be translated to covering $N = 2k_i \delta (2\theta)/S_{\bm{s}} \approx 5.5\times 10^3$ speckles before the out-of-detector-plane mismatch increases to of significant influence (1/2 of BD, the out-of-detector-plane speckle size).

\subsection{Treatment for in-detector-plane mismatch}
\label{section:spatialCorrelation}
As shown in Fig.~\ref{fig:solution}, even though the out-of-detector-plane $Q$ mismatch is well compensated by using both different wavelength and incident angles, the in-plane mismatch cannot be cancelled, we have the in-plane mismatch: 
\begin{equation}
    \mathrm{BB'} = \frac{\mathrm{OB}\sin(\eta/2)}{\sin(\theta-\eta/2)} \approx k_i \eta + k_iO(\eta^2)%%\frac{k_i \eta^2 }{2\tan\theta}.
\end{equation}
The speckle patterns from the two branches will have an offset in the direction of crossing. As $\eta \sim 10^{-4}$, the offset is to the first order invariant of scattering angle $2\theta$. For the beam parameters discussed above, $\mathrm{BB'}\approx 5.1\times 10^{-4} \mathrm{\AA^{-1}}$ is larger than in-detector-plane speckle size, and the sum of the speckle patterns will be shifted by tens of speckle sizes. As a result, the visibility analysis which calculates intensity correlation from the scattering of the two branches at the same detector location will not work.

In this case, the dynamics information regarding the sample can be extracted via the spatial intensity autocorrelation of the summed speckle patterns. Using $i, j$ to indicate the pixel $p_{i,j}$ falling into the chosen ROI on a 2D detector, and assuming there is a vertical mismatch $s$ in the speckle pattern between the two pulses. $s$ corresponds to the BB' in the reciprocal space as mentioned above. $f$ denotes the frame number recorded. Using $\Delta t$ to denote the time separation between the two pulses in a pulse pair, define $$I_f = I_{1,f}(t) + I_{2,f}(t+\Delta t),$$ the intensity correlation between pixel $p_{i,j}$ and $p_{i,j+s}$ can be estimated with the following equation:
\begin{widetext}
\begin{equation}
    A(p_{i,j}, s,\Delta t) = \frac{1}{N_f}\frac{\sum_{f=1}^{N_f} (I_{1,f}(p_{i,j},t)+I_{2,f}(p_{i,j},t+\Delta t))(I_{1,f}(p_{i,j+s},t)+I_{2,f}(p_{i,j+s},t+\Delta t)) }{(\overline{I_1(p_{i,j})}+\overline{I_2(p_{i,j})})(\overline{I_1(p_{i,j+s})}+\overline{I_2(p_{i,j+s})})}
\end{equation}
\end{widetext}
Intensity average for each pixel $p_{i,j}$ is
\begin{equation}
    \overline{I_n(p_{i,j})} = \frac{1}{N_f} \sum_{f=1}^{N_f} I_{n,f}(p_{i,j})
\end{equation}
Here $n = 1,2$ denotes the first or second pulse in a pulse pair. 

Define $r$ as the fraction of the first pulse intensity:
\begin{equation}
    r = \frac{\overline{I_1}}{\overline{I_1}+\overline{I_2}}
\end{equation}
\begin{equation}
 \begin{split}
    A(p_{i,j}, s,\Delta t) &= \frac{\overline{I(p_{i,j})I(p_{i,j+s})}}{\overline{I(p_{i,j})}\cdot\overline{I(p_{i,j+s})}} \\
    &= 1+r^2-r +  \frac{\overline{I_1(p_{i,j})I_2(p_{i,j+s})}}{\overline{I(p_{i,j})}\cdot \overline{I(p_{i,j+s})}}
    \label{eq:spatial correlation}
\end{split}   
\end{equation}
Averaging over ROI covering an iso-$Q$ range, we have:
\begin{equation}
A(Q,s,\Delta t) = \frac{1}{N_{ROI}}\sum_{i,j \in ROI(Q)}A(p_{i,j}, s,\Delta t).
\end{equation}
Here $N_{ROI}$ is the number of pixels enclosed in the ROI.
Siegert relation~\cite{goodman2015statistical} states that
\begin{equation}
    \begin{split}
    g_2(Q,\Delta t)& = \frac{\langle I_1I_{2s} \rangle }{\langle I_1\rangle \langle I_{2s}\rangle}\\
    &= \frac{1}{N_{ROI}}\sum_{i,j \in ROI(Q)} \frac{\overline{I_1(p_{i,j})I_2(p_{i,j+s})}}{\overline{I_1(p_{i,j})}\cdot \overline{I_2(p_{i,j+s})}} \\
    &=  1+\beta|f(Q,\Delta t)|^2.
\end{split}
\end{equation}
Here $I_{2s}$ indicates the speckle pattern of the second pulse shifted by $s$ in order to get aligned with that of the first pulse. 
The spatial intensity correlation encodes sample information:
\begin{equation}
\begin{split}
A(Q,s,\Delta t) &= 1+r^2-r+r(1-r)g_2(Q,\Delta t) \\
&= 1+(r-r^2)\beta|f(Q,\Delta t)|^2.
\end{split}
\end{equation}
The only additional assumption is that correlations should show negligible variation over $\delta Q$ = BB'. When $r = 0.5$, this equation describes the equal intensity case:
\begin{equation}
A(Q,s,\Delta t) = 1+\frac{1}{4}\beta|f(Q,\Delta t)|^2.
\label{eq:spatial_correlation}
\end{equation}
In conclusion, in the detector plane, as scattered photons from $Q$ and $Q+\delta Q$ fall into the same location on the detector, we need to measure coincidence of photons $\delta Q$ apart. And this can be calculated directly via the spatial correlation of the recorded 2D scattering sum, with the decorrelation between the two speckle patterns revealed in the decrease of the side band peak magnitude. 
\section{Simulation of the solution}
\begin{figure}[h!]
\centering
\includegraphics[width=\linewidth]{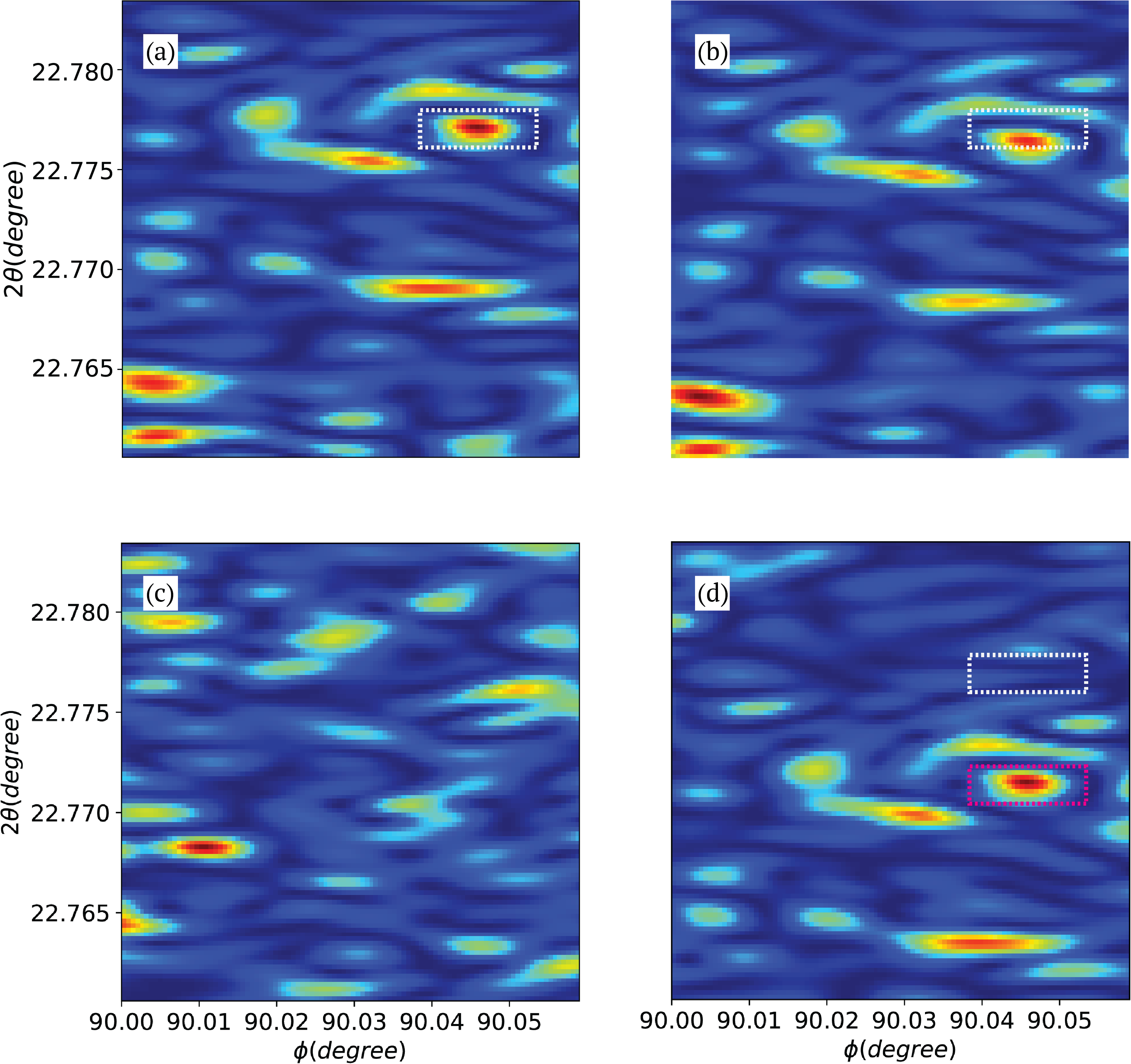}
\caption{Simulation of speckle patterns using wavelength and wavefront splitting assuming the two x-ray beams illuminating a $1.6~\mu m \times 1.6~\mu m \times 15~\mu m$ volume of random scatterers. The detector is placed at vertical scattering geometry 12.4~meter from the sample with $50~\mu m$ pixel size. (a) The speckle pattern of the beam in orange with center photon energy 10~keV. Its lower left corner corresponds to $2\theta \approx 22.76^{\circ}$ and $\phi = 90^{\circ}$ with respect to the beam denoted in orange as shown in Fig.~\ref{fig:wavefront} and~\ref{fig:wavelength}. (b) The speckle pattern of the beam in green as illustrated in Fig.~\ref{fig:wavelength} with a difference in center wavelength $\delta \lambda$ compared to the orange beam ($\delta \lambda/\lambda = 5.6 \times 10^{-5}$). (c) The speckle pattern of the green beam as illustrated in Fig.~\ref{fig:wavefront} with $\eta = 10^{-4}$ in the vertical direction. Both beams have the same center photon energy. (d) The speckle pattern after using a different photon energy for the green beam as illustrated in Fig.~\ref{fig:solution} to compensate for the out-of-detector-plane mismatch of scattering caused by the crossing angle $\eta = 10^{-4}$. Here the difference in center photon energy or wavelength $\Delta
\lambda$ satisfies $\Delta \lambda /\lambda \approx 5 \times 10^{-4}$. The white dashed boxes in (a) (c) (d) enclose the same $2\theta$ and $\phi$ range.}
\label{fig:simulation}
\end{figure}

\begin{figure}[h!]
\centering
\includegraphics[width=\linewidth]{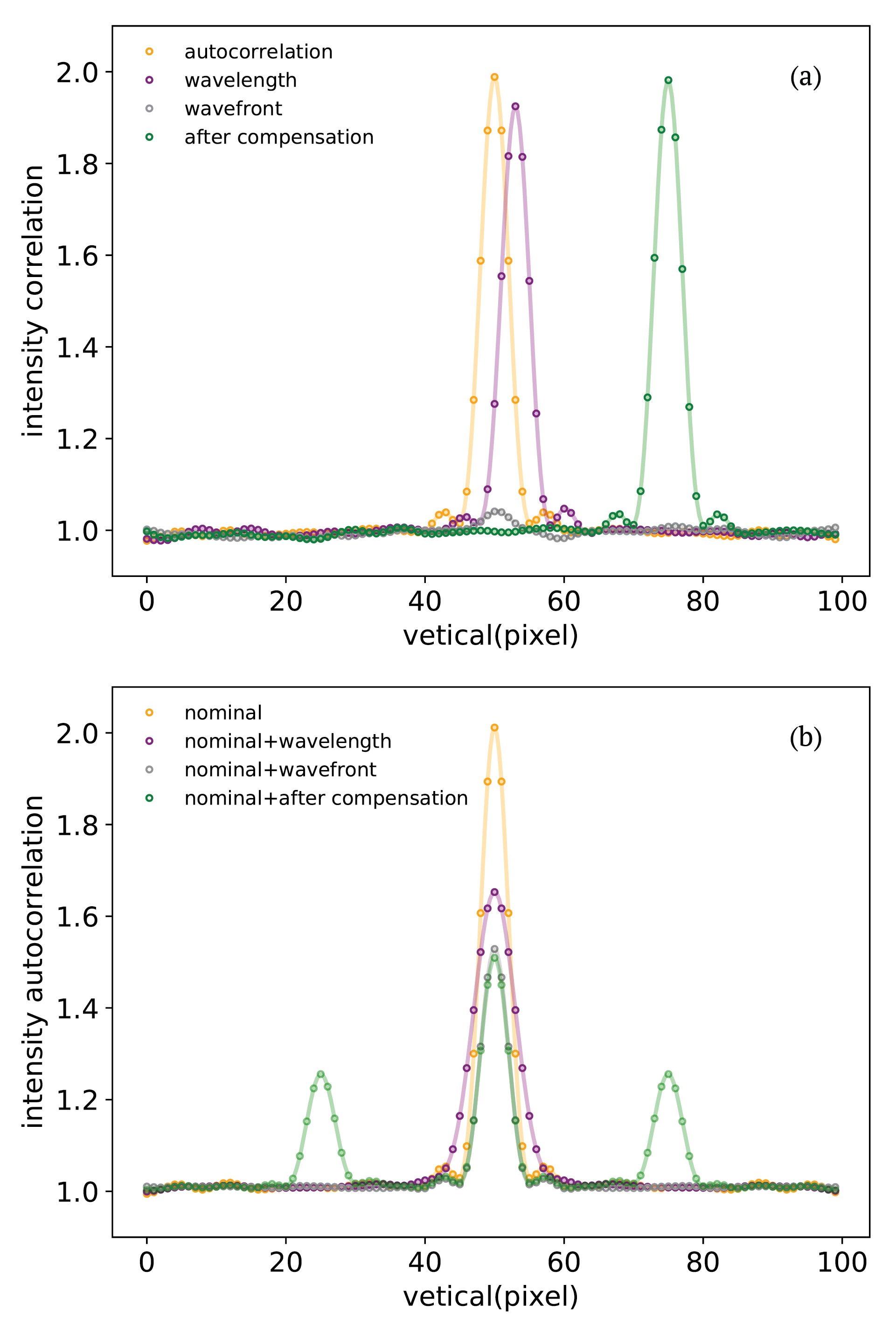}
\caption{Spatial correlations of the simulated speckle patterns in Fig~\ref{fig:simulation} in the vertical direction. (a) Orange: intensity autocorrelation of the nominal speckle pattern as shown in Fig.~\ref{fig:simulation}(a). Purple: intensity cross correlation between the nominal and the other beam from wavelength splitting (Fig.~\ref{fig:simulation}(b)). Gray: intensity cross correlation between the nominal and the other beam from wavefront splitting (Fig.~\ref{fig:simulation}(c)). Green: intensity cross correlation between the nominal and the compensated speckles (Fig.~\ref{fig:simulation}(d)).  (b) Orange: intensity autocorrelation of the nominal speckle pattern as shown in Fig~\ref{fig:simulation}(a). Purple: intensity autocorrelation of the speckle sum of the nominal and the other beam from wavelength splitting (Fig.~\ref{fig:simulation}(b)). Gray: intensity autocorrelation of the speckle sum of the nominal and the other beam from wavefront splitting (Fig.~\ref{fig:simulation}(c)). Green: intensity autocorrelation of the sum of the nominal and the compensated speckle patterns (Fig.~\ref{fig:simulation} (d)).}
\label{fig:correlation}
\end{figure}
Using the same beam parameters, we performed a simulation by calculating the coherent scattering from an illumination volume of $1.6~\mu m \times 1.6~\mu m \times 15~\mu m$ random scatterers, with 15~$\mu m$ being the sample thickness along the beam direction $\bm{k_i}$. A detector with 50 $\mu m$ pixel size was placed 12.4~meter downstream the sample in the vertical scattering geometry ($\phi = 90^{\circ}$) to oversample the speckles in the scattering. Shown in Fig.~\ref{fig:simulation}(a) is the scattering of the nominal beam denoted in orange as shown in Fig.~\ref{fig:wavefront},~\ref{fig:wavelength} and~\ref{fig:solution} with 10~keV center photon energy and beam incidence along the sample thickness direction. The lower left corner of the speckle pattern corresponds to a momentum transfer of $Q = 2 \mathrm{\AA^{-1}}$ ($2\theta = 22.76^{\circ}$) and $\phi = 90^{\circ}$. Due to the vertical scattering at high angles and the illumination dimension nearly an order of magnitude larger along the incident beam direction, the speckle size is smaller in the vertical direction on the detector. Its intensity autocorrelation in the vertical direction is plotted in orange in Fig.~\ref{fig:correlation}(a) and (b) as reference. The small side lobes are due to the non-Gaussian illumination. We plotted in Fig.~\ref{fig:solution}(b) and (c) the speckle patterns of the green beams illustrated in Fig.~\ref{fig:wavelength} and Fig.~\ref{fig:wavefront} that slightly deviate in the center wavelength or incident angle from the orange beam due to the wavelength/wavefront splitting. Using wavelength splitting, the out-of-detector-plane mismatch is a factor of 2.6 smaller as compared to the speckle size.  The in-plane mismatch component leads to the speckles shifting in the vertical direction by almost one speckle size. As displayed in Fig.~\ref{fig:solution}(b), we can still visualize shifted but similar speckles with a change in the intensity distribution. The center peak shift and value reduction in the cross correlation between this speckle pattern and the nominal one plotted in purple in Fig.~\ref{fig:correlation}(a) confirm both in- and out-of-plane mismatch from our previous calculation. This is neither optimized for visibility nor spatial intensity correlation analysis when only their speckle sum can be measured. The autocorrelation of the sum is also drawn in purple in Fig.~\ref{fig:correlation}(b). From this we can see that the shift leads to a broader center peak and contrast reduction to close to 0.6. It is important to note here that it is impractical to make the difference of the wavelengths of the two pulses $\Delta \lambda$ larger in order to fully separate the same speckle measured by the two beams, as this will require extremely small sample thickness due to the increase of the out-of-detector-plane mismatch, which is also proportional to $\Delta \lambda/\lambda$. Narrow bandwidth reflection$~\sim 10^{-5}$ will be preferred for optimizing the geometry for visibility analysis.

For Fig.~\ref{fig:solution}(c), even though $\bm{k}_i'$ (green) has a crossing angle $\eta$ with respect to that of $\bm{k}_i$, we still choose the $2\theta$ to be the scattering angle of the exit wavevector with respected to $\bm{k}_i$ as this relates to the same location on the detector. We can see that as the out-of-detector-plane speckle size is very small compared to the speckle mismatch with $2 \mathrm{BD}/\mathrm{BC} \approx 0.30$ in this case. The detector is actually sampling different speckle ellipsoids. As a result, we are not able to identify similar speckle patterns any more. Its cross correlation with the nominal speckle pattern together with the autocorrelation of their sum plotted in gray in Fig.~\ref{fig:correlation}(a) and (b) suggest that the detector cannot detect correlation anymore as it is imaging different speckles in the reciprocal space. Shown in Fig.~\ref{fig:solution}(d) is the speckle patterns after we use a different center photon energy of the green beam to compensate for the effect of the crossing angle. As mentioned, with $\Delta \lambda/\lambda \approx 5.0\times 10^{-4}$, the out-of-detector-plane mismatch can be fully compensated, this is why we can again visualize the exact same speckles, as indicated by the pink dashed box. However the in-detector-plane mismatch is even larger as the effects from the crossing angle and different wavelengths add up. The information regarding sample dynamics can be extracted from the shifted speckle sum using spatial intensity correlation analysis as mentioned in the previous section. The cross correlation of the nominal and compensated speckle patterns plotted in Fig.~\ref{fig:correlation}(a) shows that the two speckle patterns are shifted but highly correlated, and the autocorrelation of their sum is plotted in green in Fig.~\ref{fig:correlation}(b) where we can see two side lobes with correlation value equal to 1.25, which is what we calculated using $r = 0.5$ from Eq.~\ref{eq:spatial_correlation}.
\section{Discussion}
\subsection{Mitigation with long beamline}
\begin{figure}[h!]
\centering
\includegraphics[width=\linewidth]{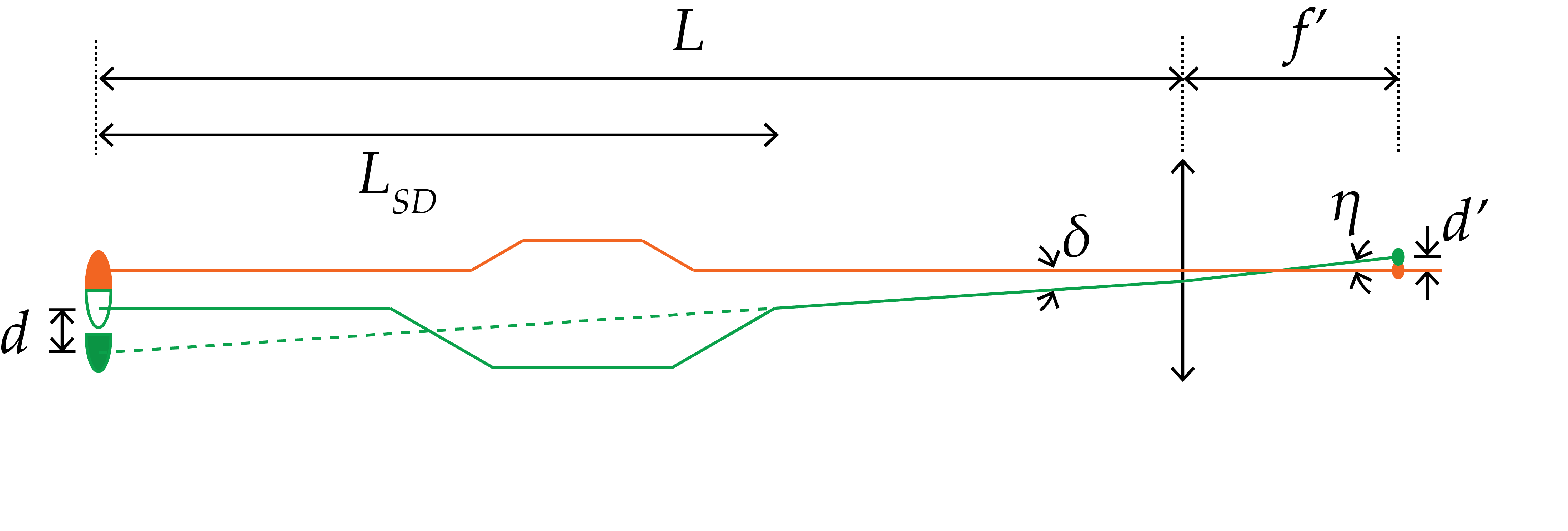}
 \caption{A general source-to-sample schematics including the split-delay optics and the focusing optics.}
\label{fig:beamline}
\end{figure}
So far we have only discussed about the scenario assuming the split-delay optics is much closer to the focusing optics compared to the distance to the source. This is the case for most current systems being deployed at the x-ray FEL facilities. In this case, the angular speckle size and the crossing angle will be always on the same order. There is a possibility to reduce the crossing angle if space allows for a very long beamline and installing the split-delay optics far upstream closer to the source. We now consider a more general split-delay optics instrument layout as shown in Fig.~\ref{fig:beamline}, with a goal of reducing the crossing angle while still maintaining a high level of beam overlap at the sample location. We assume the orange beam is on the optical axis of the lens. One can rotate the last crystal to steer the green beam path by a small angle $\delta$ to make the two beams achieve partial spatial overlap at the lens. This effectively introduces an offset vertically to the source of the green beam with respect to the original (orange) source position by the amount
\begin{equation}
    d = L_{SD}\delta.
\end{equation}
In the lens imaging system with a focal length $f$, assume $f$ is on the order of a few meters, and $L$ is on the order of a few hundred meters, then the distance between the lens and the demagnified source image $f'\approx f$. The shift of the green beam focus can be estimated by
\begin{equation}
 d' = d\frac{f'}{L} \approx \frac{L_{SD}}{L}\delta f.
\end{equation}
In order to have the focus shift much smaller than the focus size, i.e., $w_0>> d'$. Using Eq.~\ref{eq:focalspot}, enforcing that $d'$ is 10 times smaller than $w_0$, we obtain the relation of 
\begin{equation}
\delta \approx 0.13\frac{\lambda}{w}\frac{L}{L_{SD}}.
\end{equation}
Here we notice that the factor $L_{SD}/L$, with the split-delay closer to the source, demagnifies the virtual source shift.

On the other hand, following the earlier discussion as well as the schematics shown in Fig.~\ref{fig:crossing}, the crossing angle after the focusing lens can be now written as 
\begin{equation}
\begin{split}
    \eta &\approx \frac{w+d'-(L-L_{SD})\delta}{f}\\
    & = [\frac{L_{SD}}{L} - \frac{L-L_{SD}}{f}]\delta + \frac{w}{f}.
\end{split}
\end{equation}
This presents an opportunity to minimize $\eta$ by choosing $L$ and $L_{SD}$ to fulfill the relationship of:
\begin{equation}
\frac{w}{f} \approx  [\frac{L-L_{SD}}{f}-\frac{L_{SD}}{L} ]\delta.
\end{equation}
If we use the typical values of $\lambda\approx 1~\mathrm{\AA^{-1}}$ and $w\approx L\times10^{-6}$, we will arrive at
$$ L_{SD} = \frac{L}{L/13 + f/L+1} < 13 ~\mathrm{meter},$$
or the split-delay system must be unrealistically close to the source to fulfill such requirement. One could work around this potentially by working with a beam size $w$ that has been slit down. For example, if we slit down the beam by a factor of 4, such that $w \approx L \times 10^{-6}/4$, we will arrive at $L_{SD} \approx 137$ meter which is more realistic, at a cost of reduced photon flux.
\subsection{High-speed signal processing with photon coincidence measurements}
In Section~\ref{section:spatialCorrelation}, we proposed a spatial correlation analysis scheme for handling the momentum transfer mismatch in the detector plane. This bears similarity to a related concept in dynamic light scattering introduced for suppressing multiple scattering known as the 3D cross-correlation light scattering. The concept utilizes a symmetric detection setup, where the information regarding dynamics at a momentum transfer $\bm{Q}$ can be studied via the cross correlation of the signal measured separately at $\bm{Q}$ and $-\bm{Q}$~\cite{phillies1981suppression,mos1986scattered}. We also note that using two pulses of slight different wavelengths to compensate for the out-of-detector-plane momentum transfer mismatch is very similar to the two-color dynamic light scattering  experiments demonstrated in the 1990s~\cite{segre1995two}: By using two lasers with different colors at a crossing angle corresponding to their wavelength difference, it is possible to also suppress multiple scattering while retrieving the temporal fluctuations in the scattering. As the same momentum transfer is located at two different spatial locations for the two colors and the detection can be color filtered, sample information is thus also encoded in the cross correlation of the signal measured. For the above DLS experiments, thanks to the extremely high coherent flux of optical lasers, fast point detectors measuring the correlations of a speckle pair are sufficient to achieve enough signal-to-noise ratio. 

For XPCS studies of atomic scale dynamics using x-rays at FEL sources, scattering signal is typically significantly less than 0.1 photons per speckle per detector data acquisition window~\cite{hruszkewycz2012high}. The low count rate can be mitigated by the use of large area 2D pixel array detectors for simultaneously measurement of as many speckles in the scattering as possible. In our spatial correlation analysis, `correlation' signal comes from the pairs of speckles at a distance $\delta Q$ defined by the crossing angle of the two beams. In the case where detectors cannot temporally distinguish the two scattering patterns and thus only record the sum of the two patterns, the observable becomes the rate of coincidence of photons in the scattering sum separated by $\delta Q$. Instead of retrieving speckle visibility by looking at photon counting statistics, the coincidence rate can be relatively easily extracted from a 2D sensor array by employing a field programmable gate array based spatial corrector on board the x-ray detector~\cite{madden2011real}. This alleviates significantly the burden of reading out and storing the full image data. In face of the upcoming increase of the source repetition rate and multi-mega-pixels detectors, this provides an effective avenue towards taking full advantage of the various new technologies, and can render ultrafast XPCS using x-ray FEL sources an effective probe of the dynamics in complex matters.
\section{Conclusion}
In summary, we presented detailed analysis of the $Q$ space sampling in the context of split-pulse XPCS experimental concept and the current split-delay optics implementations. We provide also discussions of the practical impact based on real experimental parameters at existing x-ray FEL beamlines. We show that the out-of-detector-plane momentum-transfer mismatch of the scatterings needs first to be reduced to well below the speckle size along that direction in order to preserve the correlation between the two successive scattering patterns from the pulse pair. For the in-detector-plane speckle mismatch, which renders visibility spectroscopy infeasible, we show that dynamics can still be extracted from the summed speckle patterns by spatial intensity autocorrelation analysis. We propose a method using two pulses of different photon energies to compensate for their different incident angles in the case when beam crossing angle is in the scattering plane. These modification to the data collection and analysis protocol are critical for realizing two-pulse XPCS for the measurement of ultrafast equilibrium dynamics in complex matter.\\
\begin{acknowledgements}
The authors would like to thank helpful discussions with Matthieu Chollet, Takahiro Sato, Sanghoon Song, and Ichiro Inoue. This work is supported by the U.S. Department of Energy, Office of Science, Office of Basic Energy Sciences under Contract No. DE-AC02-76SF00515.
\end{acknowledgements}

\bibliography{reference}
\end{document}